\documentclass[journal]{IEEEtran}

\usepackage{amsmath}

\usepackage{url}
\usepackage{graphicx}  
\usepackage{float}  
\usepackage{hyperref}

\begin{document}

\title{Non-Functional Requirements in Medical Imaging}

\author{Amanda Garde Vallentin, amgv@itu.dk, 19933}
        
\markboth{IT University of Copenhagen, MSc in Software Design, research project, may 2023}
{Shell \MakeLowercase{\textit{et al.}}: Bare Demo of IEEEtran.cls for IEEE Journals}

\maketitle

\begin{abstract}
The diagnostic imaging departments are under great pressure due to a growing workload. The number of required scans is growing and there is a shortage of qualified labor. AI solutions for medical imaging applications have shown great potential. However, very few diagnostic imaging models have been approved for hospital use and even fewer are being implemented at the hospitals. The most common reason why software projects fail is poor requirement engineering, especially non-functional requirements (NFRs) can be detrimental to a project. Research shows that machine learning professionals struggle to work with NFRs and that there is a need to adapt NFR frameworks to machine learning, AI-based, software. This study uses qualitative methods to interact with key stakeholders to identify which types of NFRs are important for medical imaging applications. The study was done on a single Danish hospital and found that NFRs of type Efficiency, Accuracy, Interoperability, Reliability, Usability, Adaptability, and Fairness were important to the stakeholders. Especially Efficiency since the diagnostic imaging department is trying to spend as little time as possible on each scan. 
\end{abstract}


\section{Introduction} \label{intro}

\IEEEPARstart{N}{owadays} our healthcare system relies heavily on imaging data to diagnose and treat patients: 90\% of all healthcare data is imaging data \cite{zhou2021}. All this data must be analyzed and is in most cases done by radiologists. However, the increase in imaging data has overtaken the number of radiologists. The results can be missed findings and long turn-around times which will jeopardize the patients' safety \cite{zhou2021}. In Denmark it is especially the guarantee of treatment of cancer that takes up many of the radiologists' resources and because the investigation of most diseases today has evolved to require scans \cite{hildebrandt2022}. There is a potential for deep learning models to relieve some of the pressure at the hospitals because they, like radiologists, can learn to recognize patterns. For this project, I will focus on computer-aided detection (CADe) and diagnosis (CADx) medical imaging applications. These technologies are used to localize and classify entities of medical scans \cite{zhou2021} and have the potential to lessen workload the workload of radiologists as well as improve diagnosing of patients. 

\subsection{Current State of Medical Imaging} 
As of 2022, 521 AI-enabled medical devices have been approved by the FDA (Food and Drug Administration in the US). Out of those, 391 devices are in radiology \cite{colangelo2022}. This number includes all kinds of models and algorithms that are based on AI, consequently, we do not know how many of those are medical imaging applications. The MONAI lab (Medical Open Network for Artificial Intelligence) published a report in 2021 that highlights that very few, if any, of the developed medical imaging applications have been implemented in hospitals \cite{zhou2021}. They point towards the integration of models in the clinical workflow as one of the biggest challenges for successful implementation \cite{zhou2021}. Chan et al. emphasize that many of the radiologist’s tasks are too complex for current deep learning models, for example comparing two patient scans to detect changes \cite{chan2019}. Liu et al. highlight that most medical imaging studies do not validate their results externally or compare their performance to radiologists \cite{liu2019}. Varoquaux and Cheplygina document how research in the area is guided by data set availability rather than clinical relevance and how this will lead to diminishing returns when continuing the research in medical imaging \cite{varoquaux2022}. Given the circumstances, there seems to be a lack of focus on the domain which might jeopardize further development in the field and implementation of the models in the hospitals. 

\subsection{The Importance of Non-Functional Requirements}
When looking at traditional software projects, it is well-documented that many software development projects fail. The CHAOS report from 2015 presents that the percentage of failed and challenged projects is still high. In 2015 only 29\% of the projects in their database were considered successful \cite{chaos2015}. Many project failures can be directly linked to poor requirements gathering, analysis, and management \cite{mandal2015}. This can for example occur when users are not involved at all or only at the beginning of the process or when there is miscommunication between the client and the development team \cite{mandal2015}.

Like it is for traditional software projects, Requirement Engineering (RE) should be a key process in the development of medical imaging applications because it ensures that the developers understand the client and user needs and thereby gain domain focus in the development process.

RE is a discipline within Software Engineering (SE) where the goal is to develop requirements for a software system that describe what the system should provide and its constraints \cite[Chapter~4]{sommerville2016}. There are two types of requirements: functional and non-functional requirements (NFR). Functional requirements describe the behavior of the system, NFRs are constraints on the system \cite[Chapter~4]{sommerville2016}. 
According to Sommerville, it can be detrimental to a system if an NFR is not being met, for example, if an aircraft system does not meet reliability requirements, it is not safe and will not be used \cite[Chapter~4]{sommerville2016}. Also, the NFRs may affect the overall architecture of the system which can make them expensive and hard to meet. Therefore, it is important to put great emphasis on NFRs in the RE process because of their significant effect on the system in question.

\subsection{Non-Functional Requirements in Medical Imaging}
Habibullah et al. recently published a study where they interviewed Machine Learning (ML) professionals about their current use of NFRs. They found that most interviewees struggled with defining and measuring NFRs for ML systems and that some NFRs from the development of traditional software need to be redefined to be applicable or relevant for ML models \cite{habibullah2023}. This suggests that there are also challenges ahead for those medical imaging developers that want to work with defining NFRs because the framework has not been adapted properly to the type of software they are developing.  

The goal of this research project is to take the first step to create a framework that can help medical imaging developers implement NFRs in their application, to ensure that in the future, more medical imaging models will be implemented in the hospital to lessen the workload of the radiologists. 

\section{Methodology} 
The overall approach of the project is to follow the principles of the RE process process which consist of three activities \cite[Chapter~4]{sommerville2016}:

\begin{itemize} 
    \item Elicitation and analysis (discovering requirements by interacting with stakeholders) 
    \item Specification (converting requirements into a standard form) 
    \item Validation (checking that the requirements define what the customer wants) 
\end{itemize} 

The activities are performed iteratively to ensure that the SE team understand exactly what the client needs and they can avoid making expensive changes to the requirements of the product \cite[Chapter~4]{sommerville2016}. Each time the activities are performed, the requirements grow more specific. In this project, I will focus on the elicitation and analysis step and the specification step. 

Traditionally, these steps are performed when you have a specific design problem in mind, for example developing an imaging model that can detect liver tumors. Time will show whether this approach will result in an actual framework or something unexpected. Additionally, I have chosen to focus on the ``user'' stakeholders. Partly because of the scope of the research project but also because the MONAI lab pointed out that the biggest roadblock for medical imaging applications is to merge the models into the clinical workflow. The key users are naturally radiologists, the doctors who describe the scans but potentially also radiographers who perform the scans on the patients. 

\subsection{Qualitative Methods}
For the elicitation of requirements, I have chosen to use qualitative methods. These types of methods are ideal when the goal is to understand a new field, you are not familiar with \cite{thagaard2004}. Since I have limited knowledge about the field of diagnostic imaging and hospitals in general, these types of methods seem appropriate. Methods like interviews and observations are characterized by the researcher using herself as an instrument to extract data. This raises ethical dilemmas because it is inevitable to build a relationship with the subjects which might influence the results, and secondly, the researcher should be careful not to misuse the subjects' trust \cite{thagaard2004}. Furthermore, it is important to address the threats to the validity of the study. Below each method I will reflect upon the validity and categorize the threats as researcher's bias (negative influence of researcher's knowledge and assumptions), reactivity (influence of researcher's presence), and respondent bias (subjects not being honest or trying to please the researcher) \cite{kriukow2018}.

\subsection{Interview} \label{interview}
Interviews are a popular method for requirement elicitation and there are different types depending on the objective of the interview \cite{zowghi2005}. At the beginning of the project, I did an interview with a chief radiographer at a Danish hospital. It was a semi-structured interview with some predefined questions and themes, see interview guide in Appendix A. Since I had limited knowledge of the field at the time, I choose to do a semi-structured instead of a structured interview. The advantages of this method are that the interviewer does not need to have deep knowledge about the topic of the interview and therefore this type of interview can be held at the beginning of a project \cite{zowghi2005}. At the same time, the interviewer still has some control over the conversation since they decide the overall themes, which is not the case with a completely unstructured interview, where it is the interviewee who decides the themes \cite{thagaard2004}. However, it can be difficult for an interviewer to remain in control during the semi-structured interview because the interviewee also is allowed to bring in new themes that may or may not be relevant. 

\subsubsection{Objective of Interview}
The main objective of the interview was to gain a basic understanding of the domain diagnostic imaging. Another objective was to gain the trust of the chief radiographer and as a result, they would open the gate of the hospital and allow me to do research there. Before the interview, I send them an agenda of the themes I wanted to cover as well as some questions below each theme. On the agenda, I also expressed how many field visits I would like. I framed it as a ``Ønskescenarie'' (Wish Scenario) to emphasize that it was negotiable. During the interview, it emerged that the chief radiographer had been very involved with the current processes of buying medical imaging applications for the whole region. They had many insights about the current implementation problems of AI medical imaging solutions and the future financial and ethical challenges of using AI for diagnostic imaging. Hence the semi-structured nature of the interview allowed them to unfold themes that I had not thought of myself.
\subsubsection{Threats to Validity}
Firstly, the interview was only done with one subject, and it cannot be ruled out that another chief radiographer would have produced different data. However, the chief radiographer was a good representative of the field since they had been involved in the decisions of buying AI solutions for the whole region. I used a dictating machine to record the interview to lessen the researcher's bias. The transcribed interview can be found in Appendix B. If only notes are taken during the interview, the researcher risk making assumptions about what the subject is saying. 

\subsection{Ethnography} 
\begin{figure}
\begin {center}
\includegraphics[width=0.45\textwidth]{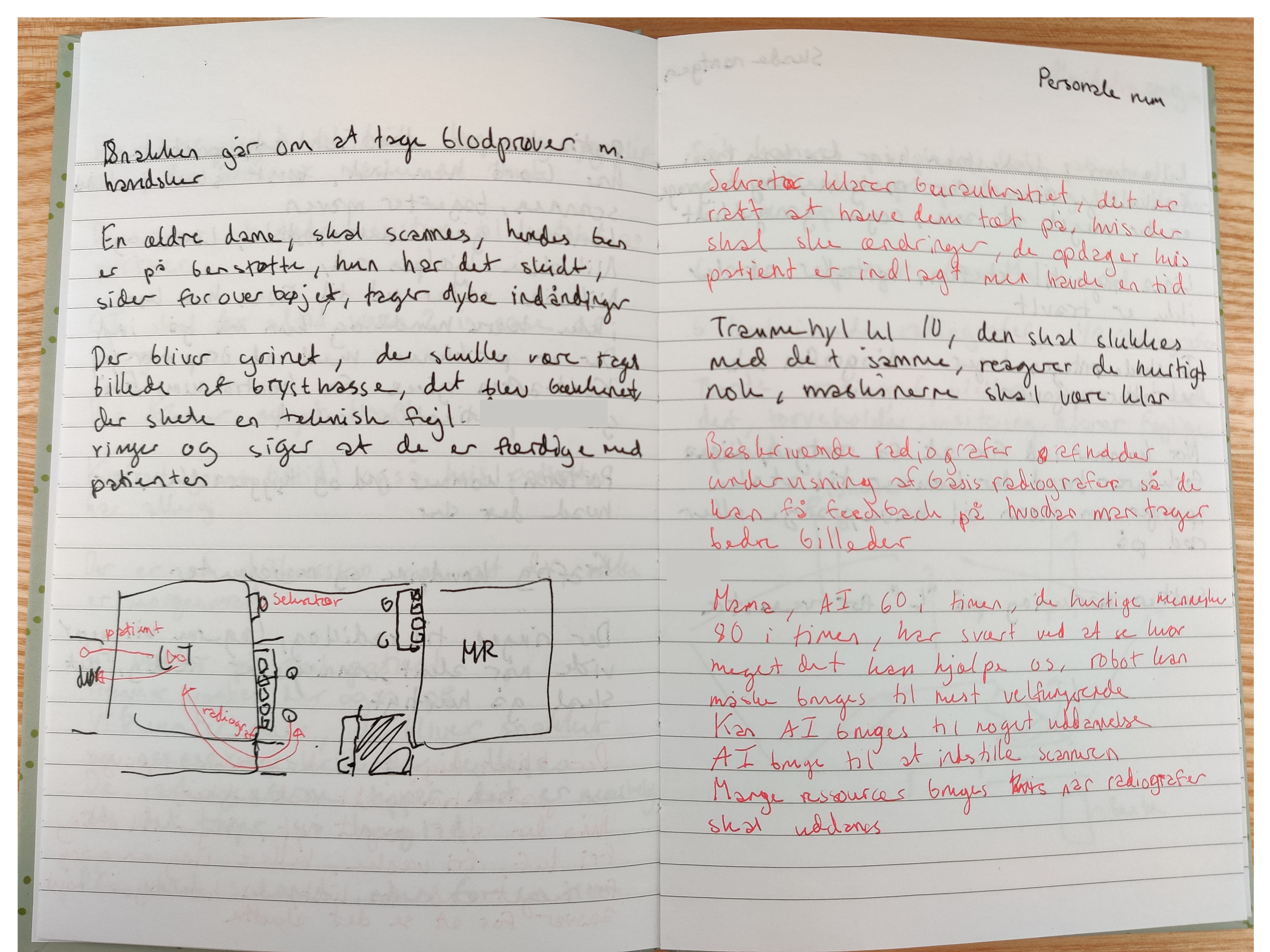}
\caption{Example of field notes from the radiographer field visit}
\label{fig:ecg}
\end {center}
\end{figure}
 Ethnographic methods seek to study people in their field over a period to gain a deeper understanding of the activities that the subject performs as well as the relationships between the different stakeholders \cite{zowghi2005}. There are degrees of involvement depending on the chosen form of observation. For example, the researcher can be a ``a fly on the wall'', or the researcher can have more of an ``apprentice role''. The downsides of these methods are that they can be very time-consuming \cite{zowghi2005}. Moreover, the subjects of study are prone to reactive bias since they can act differently because of the researcher’s presence \cite{thagaard2004}. 

\subsubsection{Objective of Ethnography}
The objective to use ethnography was to obtain an understanding of the reality of the radiographers and radiologists and thereby gaining insights in their matters of concern. This was be done by visiting a diagnostic imaging department on two different days where I first observed the radiographers and then the radiologists. Ideally, the observations would have lasted longer than just two days. I did not feel comfortable asking for more days since the diagnostic departments are under great pressure now. Although it was not ideal, I still believe I gained some valuable insights into the domain with only two days in the field. 
During my visits, I had a visible role where I also asked questions about what they were doing and why they were doing it in a certain way. Also, many of the staff at the hospital were curious about what I was doing there, and I had to explain many times what my project was about and what my background was. They were not used to having other than medical students around and naturally, we also had many conversations about AI. Therefore, these visits that were supposed to be mainly observational also turned into a bundle of smaller interviews. 

\subsubsection{Threats to Validity}
A challenge was that many of the radiologists seemed uncomfortable with me observing them when they were describing scans, which might have led them to act differently in my presence and thereby reactivity bias. I believe this is due to the nature of their work. They spend the majority of their time in front of their computers and I myself would also feel uncomfortable or watched if someone was observing how I worked on my computer. Luckily, I ended up finding one radiologist that did not seem to mind having me around and was good at explaining their process. Hence my understanding of the workflow of the radiologist is mainly based on that person. Even though this person was comfortable with being observed, they might still have been reactive bias. It was generally easier to observe the radiographers that were naturally very focused on the patients resulting in them seemingly forgetting about my presence from time to time. On both my visits, I carried a notebook where I wrote down observations and quotes from the people I talked to as well as small drawings of workflows. An example of the field notes is in Fig. 1. I could tell that this also distracted the subjects from time to time and some of them were very curious as to what I was writing down. This might have also caused respondent bias because they were aware that I was writing notes. Another aspect that should be mentioned is that most of the subjects did not have ``a choice'' regarding me observing them (as far as I am aware). Since the approval of the chief radiographer, I negotiated dates with the managing radiographer who assigned one of the radiographers and radiologists to be responsible for me. Therefore I tried to be as mindful as possible and avoided people that seemed uncomfortable with me. During data processing, I anonymized all the subjects down to their gender and tried not to include sensitive information. The demography of the subjects was mixed. There was a broad representation of different ethnic backgrounds and an overweight of women in both radiographers and radiologists. There was also a broad age representation. However, there were more older radiologists than young ones. Most of them were in their forties and fifties. There were more young radiographers than old, many of them were in their thirties.  

\subsection{Affinity Diagram}
Affinity Diagram is a way of analyzing field data from the users which reveals their matters of concern \cite{holtzblatt2004}. Before building an Affinity Diagram, all field data is condensed and written onto post-its that can be moved around. The post-its are placed on a wall in smaller groups that describe a single issue or point that is relevant to the design problem. The groups must be small, and the groups cannot be predefined but must emerge from the data. Afterwards, each group will be assigned a blue post-it that describes the point of the group in the voice of the user. For example, if there is a group of post-its describing how radiologists are busy in different ways, a blue post-it for this group might say ``I must always describe images as fast as I can''. Finally, the blue post-its are organized in areas of interest which are pink labels. Those labels are then grouped under green labels that depict whole themes \cite{holtzblatt2004}. When the Affinity Diagram is finished, the product is a hierarchical structure that describes the users’ matters of concern in different levels of detail. As a viewer, it is easy to get an overview of the users' needs. 

\subsubsection{Objective of Affinity Diagram}
The objective of the Affinity Diagram was to infer the radiographers’ and radiologists’ matters of concern from the collected data. I chose this method because it creates a visual overview of all the data and their connection to the themes. It also shows what data contributed to what themes which can make it easier to detect bias or find areas that need further research. Lastly, the data can be of different types, so it is possible to analyze both interview and observation data simultaneously. However, I had to alter the method since I had fewer data pieces than recommended and wanted to create a digital board that could easily be viewed and shared. Usually, a board is built with data from 12-16 user interviews with 50-100 notes per user \cite{holtzblatt2004}. I have been in contact with around 20 users (8 radiologists, 12 radiographers, and 1 chief radiographer) and ended up with around 100 notes on my affinity board. From my bachelor, I have experience in using the method with less data than recommended with good results, therefor I chose to use the method despite my data shortcomings. Traditionally, an Affinity Diagram is built as an analog board with paper post-its and they are built with a bigger group of people. This is mainly done to speed up the process, but also to validate the inferred themes. Depending on how much data there is, the board can get enormous which makes it inconvenient to hang somewhere in the duration of the project. 

\subsubsection{Threats to Validity}
Since I am a team of one person, I did not have the opportunity to build the diagram with team members. Hence a researcher’s bias is expected. In hindsight, I should have sought out some fellow students or researchers to help me build the Affinity Diagram, or at least have tested if someone else would have chosen the same themes over the same data. To lessen the bias I looked at the data pieces in random order to avoid pre-imposed themes. 

\section{Analysis}
\begin{figure*} \label{analysis_process}
  \includegraphics[width=\textwidth]{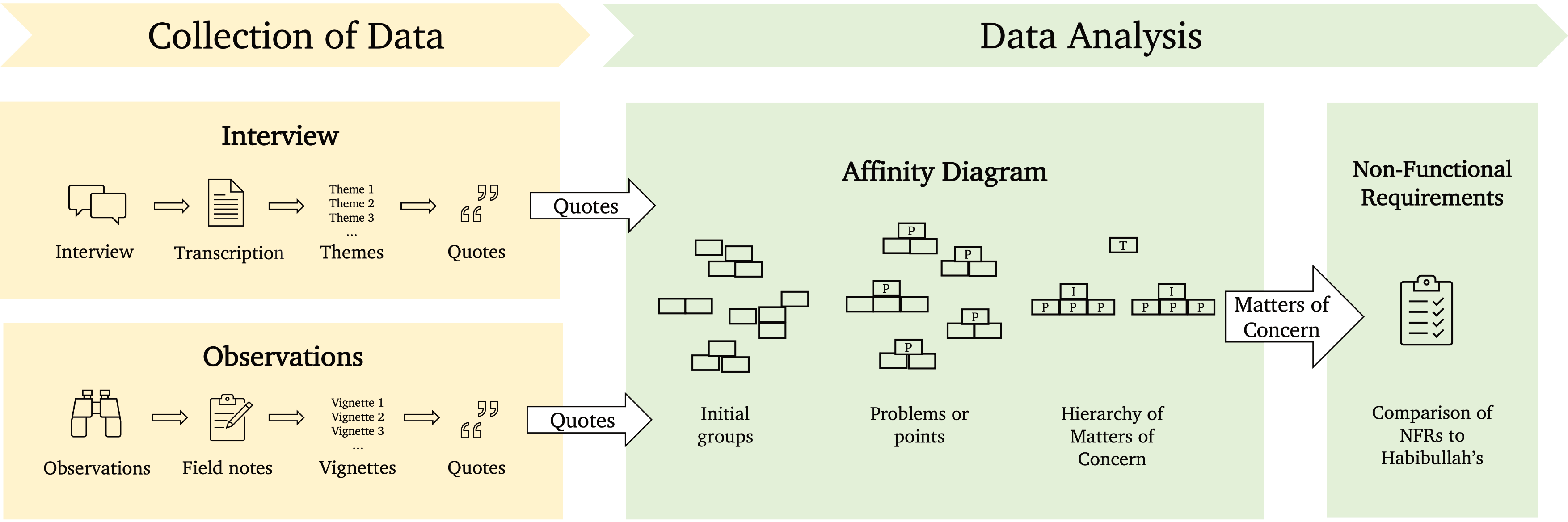}
  \caption{Analysis process. The interview data was transformed to first a transcription, then themed text, and then quotes. The observation data was transformed to field notes, then vignettes, and then quotes. All of the quotes were afterwards transformed into a hiearchy of matters of corcern through an Affinity Diagram and in the end, those were transformed into NFRs.}
\end{figure*}

The analysis process consisted of two phases: Collection of Data and Data Analysis. A visualization of the process can be found in Fig. 2. The objective of the first phase was to transform the interview and observations into smaller data pieces that could be analyzed. In the second phase, the objective was 1) transform the data into matters of concern and 2) translate matters of concern into NFRs. The process will be detailed in the following paragraphs. 

\subsection{Transform Interview into Post-It Format}
Firstly, I transcribed the interview to translate the audio file into text that could easily be analyzed. I did an initial coding of the interview where the text was grouped into themes, see Appendix C. The themes are decided by the researcher and what meaning she deducts from the different parts of the interview \cite{thagaard2004}. Coding is usually applied to identify themes and patterns across multiple interviews. As described in earlier, there was only a single interview. Therefore, I used coding to categorize the interview into parts that I believed were relevant to include in the Affinity Diagram, instead of comparing themes across interviews. Below are the identified themes of the interview:

\begin{itemize} 
    \item The limbo: Pressure to implement AI but not much to gain
    \item Cost of AI solutions
    \item Radiographers
    \item Radiologists
    \item Moving/shift of workload
    \item Developers of AI
    \item Software/AI in hospitals now
    \item Storing of AI health data
    \item Development of AI that they see potential in
    \item Who pays for AI
\end{itemize}

At last, I picked out quotes from the thematized transcription which I believed were the most descriptive of the theme. These quotes were then written on post-its in the online tool Miro so they could later enter the Affinity Diagram. This process was also susceptible to researcher bias since I was the only one coding the transcription and picking quotes. 
 
\subsection{Transform Observations into Post-It Format}
Transforming the data from the field visits were done differently from the interview since there were no audio recordings. Instead, I had my field notes and my memory to rely on. Naturally, this also introduced researcher bias. Perhaps to a greater extent than before, since there was no audio recording of the visits. When I returned from the field visits, I wrote down what I experienced either the same evening or the next morning to ensure that no important details would be forgotten. Traditionally during fieldwork, the researcher keeps account of their observations in a journal that is later used to write what is called ``thick descriptions'' \cite{thagaard2004}. Thick descriptions are extremely detailed accounts of what the researcher observed that include the cultural practices and beliefs of the subjects \cite{luhrmann2001}. Since I had limited time in the field, I simply described the situations I had observed and experienced in the hospital and did not write a journal. I chose a vignette format where I described specific situations from different locations. The vignettes can be found in Appendix D and Appendix F. If an ethnographer read the vignettes, they would probably categorize them as ``thin descriptions'' because the vignettes do not describe the meaning behind the behaviours of the radiographers and radiologists. However, the goal of the fieldwork was not to get a deep understanding of them but rather an understanding of their workflow and matters of concern. Lastly, I picked out quotes from the vignettes that were central to my observation or reflected something interesting. These quotes were also written on post-its in Miro so they could also enter the Affinity Diagram. The quotes can be found in Appendix E and Appendix G.

\subsection{Transform data into matters of concern}
When the data entered the Affinity Diagram, it was kept in three groups depending on where it came from. The groups were data from the interview with the chief radiographer, observations of radiographers, and observations of radiologists. Each group was assigned a color to ensure that when the initial groups later were broken up, you would still be able to tell where the data came from. Usually in Affinity Diagrams, the post-its have a code that describes which user the data came from. Since I only differed between three user types, I chose to give the post-its different colors so that the data source could be seen clearly by the viewer. Then the post-its were placed in random order to avoid pre-imposed themes \cite{holtzblatt2004}. When the data post-its later were grouped under blue post-its that described a point or an issue, I also choose to use different nuances of blue to indicate whether the blue post-it applied to the radiographers, the radiologists, or both. If one of the post-its could be applied to multiple categories, I placed a small colored figure in the corner to indicate they were duplicates. The different iterations of the Affinity Diagram can be found on this link:\\
\href{https://miro.com/app/board/o9J_lyx4wa4=/?share_link_id=466738437935}{https://miro.com/app/board/}\\
The resulting affinity diagram in Fig. 3 shows four different identified areas of concern which will be described in the following paragraphs.
\begin{figure*} \label{affinity}
  \includegraphics[width=\textwidth]{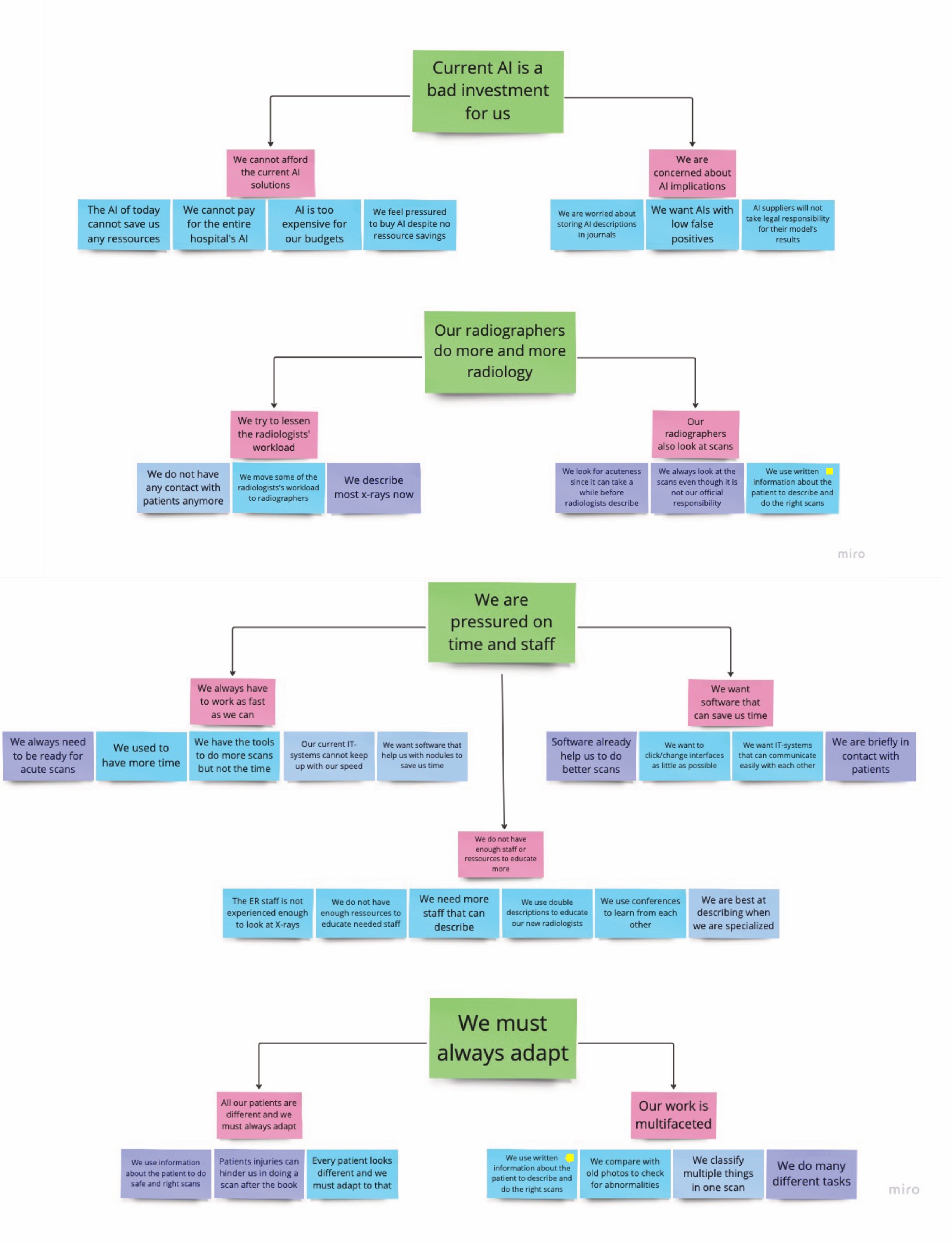}
  \caption{Resulting Affinity Diagram}
\end{figure*}

\subsubsection{Current AI is a bad investment for us}
This matter of concern stems from two issues that the hospitals are facing right now when they want to buy AI solutions for diagnostic imaging. The first issue is that they cannot afford the current solutions. If they invest in an AI solution, they need to know, that it can save them resources. The resources can be in the form of staff; maybe they can settle with 8 radiologists instead of 7, or it can be in saved time so that current radiologists can describe images faster. However, the results, they know about, show that there can only be saved minimal resources and that the investments have been loss-making transactions meaning that the cost of using AI was much larger than the saved resources. Another problem is that the hospitals lack a fair way of financing the models. Right now, the diagnostic imaging department is told that they need to pay for the models since the models work on their scans instead of for example spreading the cost evenly among all departments. The second issue is that they are concerned about what implications the AI models will have in the future for them. For example, can radiologists get sued for not trusting an AI prediction if it later turns out that the patient got sick? Also, none of the AI suppliers will take legal responsibility for their models' results, so an AI prediction always needs to be held up against a real radiologist's opinion. This is also why it is difficult to avoid the AI models being a loss-making investment because they are not legally allowed to replace radiologists. Also, it is important for the hospital that the AI solutions have a low level of false positives since these types of results potentially can make the hospital spend unnecessary resources on patients that are not sick. 

\subsubsection{Our radiographers do more and more radiology}
This matter of concern stems from the matter of concern below; they are pressured on time and staff. A solution they have been working towards in the past years is to move some of the workloads of the radiologists to the radiographers. It is easier to hire radiographers than radiologists. This means that most X-rays are not described by ``describing radiographers'' who have received special education to learn how to describe those scans properly. The ultrasounds will also be done by specialized radiographers in the future. An interesting thing that I noticed during my field visits was that the radiographers also analyzed the scans and told me about what they could see was wrong with the patient. It turns out that they also look for abnormalities in the patient to detect if there is an acute issue that needs to be looked at right away.

\subsubsection{We are pressured on time and staff}
Another major matter of concern is that both the radiologist and radiographers are understaffed while the number of required scans for patients are increasing. This also means that they have fewer resources to help educate new radiographers and radiologists which creates a negative reinforcement loop. The radiologists express that their systems are not geared to handle as many scans as they do now and that they also avoid using extra aids. For example, they have access to a program that shows the scans in better resolutions, but most of them do not use it because it takes too long to open the programs. Therefore, they request that the software they will buy in the future will save them valuable time. They want software where they do not have to click around in many different interfaces but instead have as much information as possible displayed in one interface. 

\subsubsection{We must always adapt}
The last identified matter of concern is that the work of both radiographers and radiologists is very challenging because there are many factors that needs to be considered when they scan and describe a patient. They must always adapt to the special circumstances of exactly that patient. For example, a patient can have an injury that prevents the radiographers from doing a scan after the book. Or maybe the patient has an old injury that still shows up on the scan but does not need to be treated. In general, every patient looks different which also makes it challenging to know what is abnormal and dangerous and what is not.

\subsection{Translate Matters of Concern into Non-Functional Requirements}
The following NFRs can be identified from the analysis:\\

\begin{enumerate}
\item The application must make the overall process of scanning or describing a patient faster by XX\% 
\item The AI model must have a false positive percentage under XX\%
\item The application must only take XX seconds to open and run
\item All important information must be shown in the same user interface
\item The model must be able to run on many different patients and types of scans
\end {enumerate}

In table 5 in Habibullah et al.’s article about NFRs for Machine Learning \cite{habibullah2023}, several NFRs for machine learning systems are identified and defined. Their list can be compared to the identified NFRs of this project to translate them into more general NFRs that medical imaging application developers should design for.

\subsubsection{Efficiency}
Habibullah defines efficiency as ``The ability to accomplish something with minimal time and effort'' which both the first and the second requirement is describing because the hospitals need to save resources to be willing to invest. Additionally, using the application should take up as little time as possible for the already time deficient radiographers and radiologists. Hence this general NFR relates to the first and third requirement. 

\subsubsection{Accuracy}
Habibullah defines accuracy as ``The number of correctly predicted data points out of all the data points''. The radiologists of course want predictions from the model that are as accurate as possible. However, it is especially important to them that the predictions have a low amount of false positives because that can potentially mean wasted resources on healthy patients. Consequently the accuracy NFR should be more detailed and specifying levels of sensitivity and specificity. Thus this general NFR relates to the second requirement.  

\subsubsection{Interoperability}
Habibullah defines interoperability as ``The ability for two systems to communicate effectively''. Possibly the application could be integrated with the current IT systems as an ``add in'' to ensure that the user will spend as little time as possible on opening up and running the application. Therefore this general NFR could apply to the first and third requirement. 

\subsubsection{Reliability}
Habibullah defines reliability as ``The probability of the software performing without failure for a specific number of uses or amount of time''. The application needs to have a reliable runtime and should be able to handle the big imaging data seamlessly. If the application is not reliable, the radiologist will not use it since they cannot afford to waste any time. Again, this general NFR also applies to the first and third requirement.

\subsubsection{Usability}
Habibullah defines usability as ``How effectively users can learn and use a system''. From my data, the radiologists see usability as using as few clicks as possible. They like to have all essential information they need in one interface. Hence this general NFR relates to the fourth requirement.

\subsubsection{Adaptability}
Habibullah defines adaptability as ``The ability of a system to work well in different but related contexts''. In the case of the radiologists, it could mean that the models would also work on scans that are abnormal. For example there could be medical equipment in the scan like tubes or the scan could be from an abnormal angle due to a patient's injury. Thus this general NFR applies to the fifth requirement. 

\subsubsection{Fairness}
Habibullah defines fairness as ``The ability of a system to operate in a fair and unbiased manner''. This is also related to adaptability above. The application should be able to work on all types of patients. Therefore this general NFR also applies to the fifth requirement. 

\section{Discussion}
\subsection{Evaluation of Results}
From performing the two first key activities in the RE process, elicitation, and specification, it was possible to identify five different non-functional requirements that can be linked to seven different general NFRs. However, the five identified NFRs are too broad or non-specific for developers to work with. The next steps to specify them further would be to validate the findings with the stakeholders and generally do more iterations of the RE process. An apparent limitation of the result is that the data only comes from a single Danish hospital. Further research is needed to confirm or deny whether the identified NFRs also apply to other diagnostic imaging departments in Denmark or other countries.
\subsection{Weaknesses in the Analysis}
\begin{figure*} \label{coding_board}
  \includegraphics[width=\textwidth]{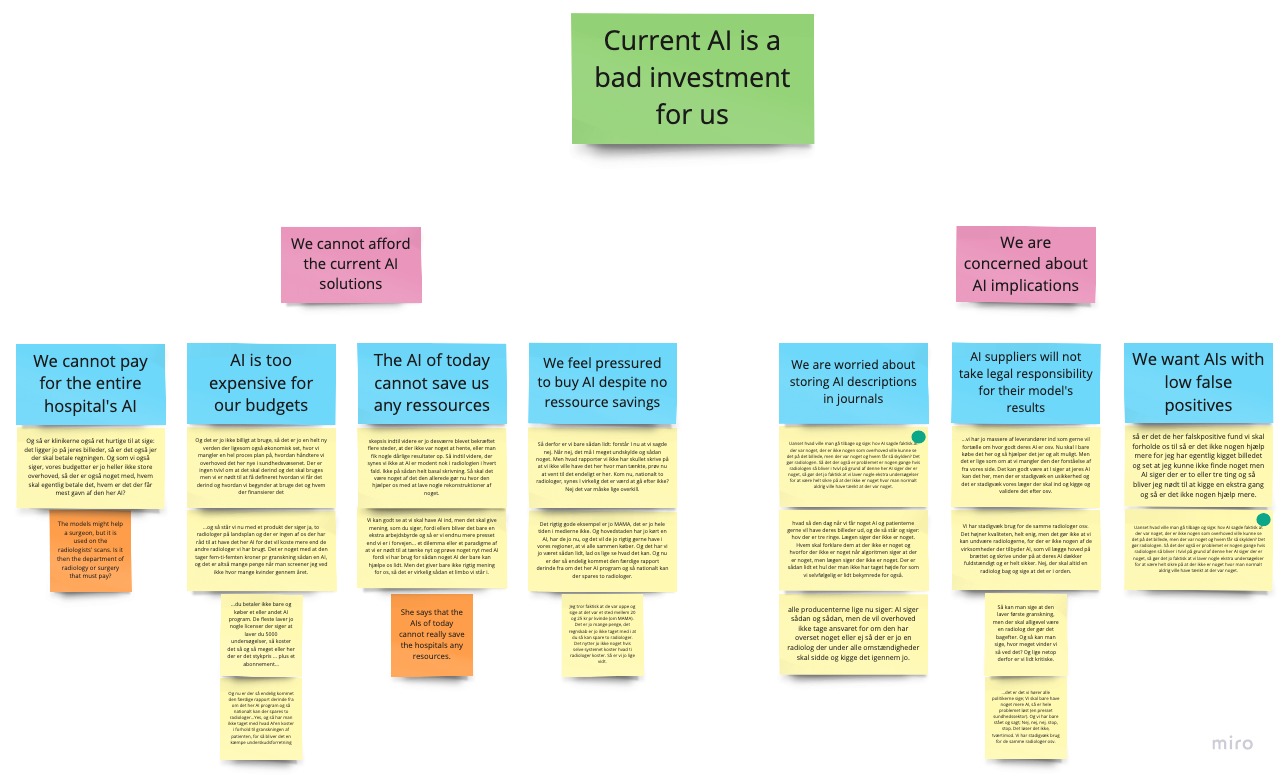}
  \caption{Excerpt of Affinity Diagram of the matter of concern ``Current AI is a bad investment for us''}
\end{figure*}
As was mentioned in the Methodology paragraph, the Affinity Diagram can reveal weak points in the analysis since you can see what data contributed to what themes. Generally, it is ideal if each blue theme post-it has data from different users to make sure that the themes are representative of all the users or a group of users. If we look at the green high-level matter of concern ``Current AI is a bad investment for us'' and follow the hierarchy down to the blue post-its, it appears that most of the yellowish data post-its have the same sand color, which means that they come from the interview with the chief radiographer, see Fig. 4. It could be argued that the credibility of those matters of concern is lower than the others. At the same time, it could be argued that it is fair to give credibility to what was said in the interview since the subject had a managing position and therefore is a representative of the diagnostic imaging departments.
\subsection{Resource Savings VS Performance}
A major finding in this study was also that there seems to be a significant difference in what the hospitals value in medical imaging AI applications and what the developers value in medical imaging AI applications. The hospitals see it as a way to save resources and can actually not afford to invest in the solutions unless they see a return on investment. Contrary to the hospitals, the developers focus on creating models that perform better than radiologists \cite{liu2019}. This is a rather unfortunate difference, and it makes me wonder if the developers are aware that their models cannot legally replace radiologists. When a radiologist still needs to check every single prediction and decide whether they believe it or not, naturally they also need to analyze the scan. Then it is difficult to imagine any amount of time saved and then it does not pay for the hospitals. These findings could be viewed as a symptom of the lack of focus on the domain that medical imaging application developers may have. However, it is also important to raise the question of whether these types of developers have the competencies or tools to understand the domain. Traditionally, these types of developers come from a data science background and are experts in analyzing data, finding patterns with the help of statistics and math, and now also machine learning. Yet they do not know much about software engineering. For example, I had to explain to my research group, PURRLab, which researches in medical imaging, what requirement engineering and non-functional requirements are. This raises another question of who exactly should oversee the requirement engineering process and who will benefit from my study. Further research is needed to determine who should undertake requirement engineering tasks for medical imaging applications.

\section{Conclusion}
By using qualitative methods to understand the matters of concern of radiographers and radiologists, this study identified general non-functional requirements. These may be used by medical imaging application developers to ensure that more applications are implemented in the hospitals and relieve some pressure on the diagnostic imaging departments. The results should be validated in further research since the data only came from a single hospital and since the analysis may be subject to researcher bias since only one person contributed to the analysis. Further research is also needed to detail the requirements.

\section{Github}
Link to github repository:\\
\href{https://github.itu.dk/amgv/NFRs-in-Medical-Imaging.git}{https://github.itu.dk/amgv/NFRs-in-Medical-Imaging.git}\\
A CSV file with all the data used in the analysis will be uploaded to the repository in the end of May 2023. In the repository, there are also notes from supervision meetings. 

\bibliographystyle{abbrv}
\bibliography{references.bib}    


\end{document}